\documentclass{article}
\usepackage{chicago}

\newcommand{\iec}{\mbox{i.\,e.\,}}

\newcommand{\egc}{\mbox{e.\,g.\,}}

\newcommand{\vctr}[1]{\ensuremath{\mathbf{ #1 }}}

\newcommand{\dr}[1]{\ensuremath{\mathrm{d} #1\,}}

\newcommand{\ddt}{\ensuremath{\frac{\dr{}}{\dr{t}}}}

\newcommand{\dbdbd}[2]{\ensuremath{\frac{\mathrm{d}^2{#1}}{\dr{#2}^2}}}

\newcommand{\be}{\begin{equation}}
\newcommand{\ee}{\end{equation}}
\newcommand{\e}[1]{\mathrm{e}^{#1}}

\begin{document}

\title{More problems for Newtonian cosmology}
\author{David Wallace}
\maketitle
\begin{abstract}
I point out a radical indeterminism in potential-based formulations of Newtonian gravity once we drop the condition that the potential vanishes at infinity (as is necessary, and indeed celebrated, in cosmological applications). This indeterminism, which is well known in theoretical cosmology but has received little attention in foundational discussions, can be removed only by specifying boundary conditions at all instants of time, which undermines the theory's claim to be fully cosmological, \iec, to apply to the Universe as a whole. A recent alternative formulation of Newtonian gravity due to Saunders (\emph{Philosophy of Science} 80 (2013) pp.22-48) provides a conceptually satisfactory cosmology but fails to reproduce the Newtonian limit of general relativity in homogenous but anisotropic universes. I conclude that Newtonian gravity lacks a fully satisfactory cosmological formulation.
\end{abstract}

\section{Introduction}

Newtonian gravity in its original force-based formulation is defined for discretely many particles by
\be \label{newton-sum}
\ddot{x}^i_n(t) = \sum_{m\neq n}G m_m \frac{(x^i_m(t)-x^i_n(t))}{|x_n(t)-x_m(t)|^3}
\ee
($m_n$ is the mass of the $n$th particle and $x_n$ is its position vector at time $t$)
or in the continuum limit by
\be \label{newton-integral}
\dot{v}^i(x,t)=\int \dr{x'}^3 G \rho(x',t) \frac{x'^i-x^i}{|x'-x|^3}
\ee
where $\rho$ is the mass density function, $v(x,t)$ is the velocity of a test particle at spacetime point $x,t$, and $\dot{v}(x,t)$ is the acceleration of that test particle, \iec the time derivative along the particle's worldline.\footnote{Throughout this paper, Roman indices range from 1 to 3, and by definition $X_i=X^i$. (I adopt no abstract index notation and intend Roman-indexed objects to be the components of vectors and tensors in standard Cartesian coordinates.)}

As has long been known (see \citeN{nortoncosmologicalwoes} for details of the history), this theory becomes ill-defined in cosmological applications, where the distribution of particles is homogeneous (\iec, if the mass density on large scales is asymptotically constant). For in that circumstance, (\ref{newton-sum}--\ref{newton-integral}) become conditionally convergent (the result of performing the sum depends on the order in which it is carried out) and thus ill-defined. 

The point can be illustrated by elementary means (again, as has long been known). Consider a test particle a distance $r$ from the central point $O$ of a sphere of matter of uniform density $\rho$. The integral (\ref{newton-integral}) can be split into two parts: one part that integrates over the matter in the smaller sphere of radius $r$ centred on $O$, and another part that sums over a series of spherical shells of infinitesimal thickness, also centred on $O$ and with radius $R>r$. It has been known since Newton that the force on a test particle external to a uniform sphere of matter is the same as would be the case if that matter were concentrated at the centre of the sphere, and that the force on a test particle inside a uniform spherical shell is zero. So the second part of that integral is zero, and the total acceleration of the particle is
\be \label{naive}
\dot{v}^i(r) = -\frac{4}{3}\pi G \rho r \hat{r}^i= -\frac{4}{3}\pi G \rho r^i.
\ee
Allowing the radius of the sphere to increase without limit makes no difference to this argument, suggesting that (\ref{naive}) is also the correct acceleration for a test particle in an infinite medium of density $\rho$. But in such a medium, the idea of a `centre' becomes ill-defined, as the matter can be decomposed into concentric spheres around an arbitrary centre.

As has been repeatedly observed, however\footnote{See, \egc, \citeN{mccrea1934newtonian}, \citeN{narlikar1963newtonian}, \citeN{davidson1973newtonian}, \citeN{evans1974newtonian}, \citeN{malament1995newtonian}, \citeN{nortonnewtoncosmology}, \citeN{ellis1997newtonian}.}, this indeterminacy in the absolute acceleration has no actually observable consequences. For what is observable is (at most) the relative acceleration of two test particles, not the absolute acceleration relative to an unobservable background. And given two test particles with vector positions $x^i_1$, $x^i_2$, the relative acceleration is
\be \label{relac}
\dot{v}^i(x_1)-\dot{v}^i(x_2) = -\frac{4}{3}\pi G \rho (x_1^i-x_2^i)
\ee
which depends on the separation of the particles but not on the location of the origin.

This suggests that in Newtonian cosmology properly formulated, only relative and not absolute accelerations matter --- or put another way, that the split between inertial motion and acceleration under gravity is arbitrary. The most straightforward way to see this mathematically is to shift from the force-based to the potential-based formulation of Newtonian physics, in which the dynamics are given by
\be 
\dot{v}^i(x,t)+ \nabla_i \Phi(x,t)=0;\,\,\,\,\,\,\,\, \nabla^2 \Phi(x,t) = 4 \pi G \rho(x,t)
\ee
(the latter being \emph{Poisson's equation}).
These equations are invariant under the transformations
\be
(x^i,t) \rightarrow (x'^i,t)=(x^i + a^i(t),t)
\ee
\be \label{potential-transform}
\Phi \rightarrow \Phi';\,\, \Phi'(x',t)=\Phi(x,t)-\ddot{a}_i(t) x^i+ V(t).
\ee
where the $a^i(t)$ and $V(t)$ are arbitrary smooth functions of time. These are arbitrary time-dependent spatial translations; the transformation law for $\Phi$ generates a time-dependent but spatially independent acceleration that compensates for the effect of the translation. 

In normal (\iec, non-cosmological) applications of Newtonian gravity, we impose a boundary condition
\be \label{bc}
\lim_{|x|\rightarrow  \infty} \Phi(x,t)=0
\ee
which serves to eliminate the gauge freedom in the equation for $\Phi$, and to restrict the arbitrary translations $a^i(t)$ to include only time-independent translations and velocity boosts. But if we drop this assumption then Newtonian gravity becomes a theory in which only relative acceleration is well-defined, and absolute acceleration is pure gauge.

In this form, the theory lends itself naturally to a geometric interpretation, according to which the inertial structure of spacetime is not the absolute structure of Galilean spacetime~\cite{anderson1967,steinnewton,earman1970,friedmanstructure} but determined locally by the matter distribution. This move is often accompanied by a differential-geometric reformulation of the theory
--- so-called \emph{Newton-Cartan theory} --- to replace the potential with a nonflat affine connection  (see, \egc, \citeN{malamentgravitybook} and references therein). \citeN{knoxequivalence} argues that even in its standard potential formulation Newtonian gravity is already a theory of dynamically-determined inertial structure; \citeN{SaundersPrincipia} goes further and argues that inertial structure can be entirely eliminated in Newtonian physics; in \citeN{wallacenewton} I argue (from the currently-unfashionable coordinate-transform route to defining physical theories; cf \citeN{wallacecoordinates}) that the transformation law (\ref{potential-transform}) for $\Phi$ means that it is already a connection. 

To see this further, consider a smooth distribution of test particles with velocities $v^i(x,t)$. The relative acceleration of infinitesimally close test particles is given by 
\be \label{relac-potential}
\nabla_j \dot{v}^i(x,t) = \nabla_j\nabla^i \Phi(x,t)
\ee
and from a geometric perspective, this is a geodesic deviation equation and identifies the symmetric matrix $\Omega_i^j=\nabla_i \nabla^j \Phi$ as the (nontrivial part of the) spacetime curvature. Poisson's equation is now
\be 
\Omega^i_i = 4 \pi G \rho
\ee
which may be thought of as a nonrelativistic version of the Einstein field equation $R_{\mu\nu}=8 \pi G T_{\mu\nu}$.

However, neither reconceptualising Newtonian gravity geometrically, nor reformulating it differential-geometrically, is required to apply it to cosmology: the simple potential-based form is already suitable (and the reformulations do not change the conclusions of this paper). In particular, the relative acceleration law (\ref{relac}) corresponds to a potential
\be \label{isopotential}
\Phi(x)=\frac{2}{3}\pi G \rho |x|^2 + b_ix^i + V
\ee
which also satisfies $\nabla^2 \Phi=4\pi G \rho$ for spatially constant $\rho$. Translations now just serve to change the linear term in (\ref{isopotential}), which is equivalent to changing the centre around which the quadratic is defined, but have no effect on the relative accelerations. 

So this seems to put Newtonian cosmology on a very satisfactory footing. And indeed, this approach to cosmology was developed by Heckmann and Schucking~\citeyear{heckmannschucking1,heckmannschucking2}, brought to wider attention by \citeN{szekeres1977} and is now the standard framework for cosmology where relativistic effects may be neglected (see, \egc, \citeN{ellisreview} for a comparison of relativistic and nonrelativistic cosmology); it also seems to have received widespread consensus in philosophy of physics (see, \egc, \citeN{malament1995newtonian}, \citeN{nortonnewtoncosmology},  \citeN{pooleyhandbook}, \citeN{knoxequivalence}).

All seems well. The only trouble is: that boundary condition on $\Phi$ was there for a reason.

\section{Non-uniqueness of solutions to Poisson's equation}

Let's review the normal route towards establishing equivalence of force-based and potential-based formulations of gravity (or indeed electrostatics). Given two solutions to Poisson's equation $\Phi_1,\Phi_2$, it follows that their difference $(\Phi_2-\Phi_1)$ satisfies \emph{Laplace's equation},
\be 
\nabla^2 (\Phi_2-\Phi_1) = 0.
\ee
The operator $\nabla^2$ is rotationally invariant and so elementary functional analysis (see, \egc, \citeN[p.95]{jackson} for details) tells us that the most general solution to Laplace's equation defined over all space is 
\be \label{gen-sol}
F(x) = \sum_{l=0}^\infty \sum_{m=-l}^l A_l^m |x|^l Y^l_m(\hat{x})
\ee
for arbitrary constants $A_l^m$ (where the $Y^l_m$ are spherical harmonics). So if $\Phi_1$ and $\Phi_2$ also satisfy the boundary condition (\ref{bc}) (so that $(\Phi_2-\Phi_1)$ tends to zero at spatial infinity), we must have $\Phi_2-\Phi_1=0$ everywhere, and so Poisson's equation has a unique solution. When we also observe that
\be 
K(x)= G \frac{1}{|x|}
\ee
satisfies
\be 
\nabla^2 K(x) = 4 \pi \delta(x)
\ee
(where $\delta$ is the Dirac delta function), we can write that general solution in closed form as
\be 
\Phi(x) = \int \dr{x'}\rho(x')K(x-x')=\int \dr{x'} G \frac{\rho(x')}{|x-x'|}
\ee
and, upon differentiation, recover Newton's force law (\ref{newton-integral}) (and thence (\ref{newton-sum}), if we specialise to a mass density that is a sum of delta functions).

But now suppose that the boundary condition is dropped. (It's not optional to drop it in cosmology: if $\rho$ is constant, no solution to Poisson's equation will satisfy it.) Then the solution of Poisson's equation is indeterminate up to a quite arbitrary solution of Laplace's equation. Some such solutions are harmless: the $l=0$ term in (\ref{gen-sol}) just corresponds to adding a constant to the potential; the $l=1$ terms to adding the linear terms we have already considered. But adding a term like
\be 
\Delta \Phi(x,y,z,t) = \Lambda(t)(2z^2-x^2-y^2)
\ee
(for arbitrary smooth $\Lambda$)
has observable physical consequences: it changes the relative accelerations. So once the boundary condition is dropped, Newtonian gravity becomes radically underdetermined.

To spell this out further, consider again the matrix $\Omega^i_j=\nabla_j\nabla^i \Phi$ responsible for generating geodesic deviation. We can decompose this matrix into its trace and traceless parts:
\be 
\Omega=\Omega^i_i;\,\,\,\,E^i_j = \Omega^i_j - \frac{1}{3}\delta^i_j \Omega.
\ee
Recall that the matrix gets physical meaning through geodesic deviation, \iec the relative acceleration of infinitesimally close test particles. Given a swarm of infinitesimally close such particles initially comoving, $\Omega$, the scalar curvature, generates uniform contraction of the swarm. $E^i_j$, the Newtonian \emph{Weyl tensor}, generates volume-preserving shear transformations.

$\Omega$ is fixed locally by Poisson's equation, $\Omega=4 \pi G \rho$. But absent a boundary condition, $E^i_j$ is vastly underdetermined $\rho$. Given a dynamically valid $E^i_j$, we can replace it with $E^i_j+\nabla_j \nabla^i \Psi$, where $\Psi$ is a quite arbitrary --- and very possibly time-dependent --- solution of Poisson's equation.

Something similar occurs in Maxwellian electrodynamics, where Gauss's law $\nabla\cdot \vctr{E}=4 \pi \rho_q$ couples the divergence of the electric field $\vctr{E}$ to the charge density $\rho_q$ and likewise needs boundary conditions for a unique solution. There, however, it suffices to fix that boundary condition at a single time: because the rate of change of $\vctr{E}$ is determined by the magnetic field $\vctr{B}$ and the current distribution $\vctr{J}$ through
\be \label{maxwell}
\dot{\vctr{E}}- \nabla \times \vctr{B} = 4 \pi \vctr{J}
\ee
we can dynamically determine \vctr{E} at later times from its initial value. (An equivalent way to say this is that $\nabla \cdot \vctr{E}-4 \pi \rho_q$ is a conserved quantity in electrodynamics, so that Gauss's law need be imposed only once.) So in electrodynamics, the boundary condition can plausibly be thought of just as a contingent initial condition. In Newtonian gravity it needs to be specified --- and can be specified arbitrarily --- at all times, not just the initial time.

Full general relativity also follows the electrodynamic model. There, the Weyl tensor has ten components: five `electric' components corresponding to the $E^i_j$ of Newtonian gravity, and another five `magnetic' components. And the two sets are related by analogues of (\ref{maxwell}). It is the vanishing of the magnetic components in the nonrelativistic limit that renders Newtonian cosmology indeterministic. (See \citeN{ellis1997newtonian} for further discussion.)

Now, in some cases there are `natural' choices of boundary conditions (for whatever that is worth). In the special case of a homogeneous and isotropic universe (as in the simplest cosmologies), it is natural to require that the relative accelerations are likewise homogeneous and isotropic. With this additional stipulation --- which amounts to the condition that the $E^i_j$ vanish --- we can recover the potential (\ref{isopotential}). But this condition must be understood as additional to the dynamical equations and, more importantly, does not extend to the (physically more realistic) cases where exact isotropy and homogeneity fail.

For instance, there are well-known generalisations of isotropic and homogeneous cosmology to the homogeneous and \emph{non}-isotropic case, and it is of interest to extend these generalisations to the Newtonian regime (see, \egc, \citeN{narlikar1963newtonian}, \citeN{szekeres1977}). But now the physical motivation for eliminating the $E^i_j$ is obscure.

We may also wish to consider small deviations from homogeneity. In cosmological perturbation theory, for instance, it is standard to decompose the matter distribution as
\be 
\rho(x,t)=\bar{\rho}(t) + \epsilon \delta \rho(x,t) 
\ee
with $\epsilon$ small, and decompose $\Phi$ likewise into a sum of the homogeneous potential and a perturbation,
\be 
\Phi(x,t)=\frac{2}{3}\pi G \bar{\rho}(t)|x|^2 + \epsilon \delta \Phi(x,t),
\ee
so that $\delta \Phi$ satisfies
\be 
\nabla^2 \delta \Phi = 4 \pi G \delta \rho.
\ee
Normally $\delta \rho$ is decomposed into Fourier modes,
\be 
\delta \rho_k = \exp(i k_ix^i),
\ee
and we take
\be \label{perturb}
\delta_k \Phi = -\frac{4\pi G}{k_ik^i}\exp(i k_ix^i).
\ee
But nothing stops us adding a quite arbitrary solution of Laplace's equation to (\ref{perturb}), so that the perturbation deviates wildly away from this form as $\epsilon$ is increased.

This is a recognised problem in theoretical cosmology (for discussion see, \egc, \citeN{heckmannschucking1}, \citeN{ellis1997newtonian}, \citeN{szekeres2000post}, \citeN{rainsford2001anisotropic}), where it is normally solved by some combination of physical good sense and appeal to the low-energy limit of general relativity. But neither is really satisfactory if we want to understand Newtonian physics as a well-defined cosmological theory in its own right.

\section{Rehabilitating the force-based approach}

Let's review how we got into this mess. The usual force-based formulation of Newtonian mechanics stipulates a unique absolute acceleration for particles. We were led to reformulate it in terms of relative acceleration, and did so via the potential formalism, in which we have the freedom to add linear terms to the potential that leave the relative accelerations unchanged. But this  reformulation gave us \emph{too much} freedom, in that it also permitted us to add terms to the potential which changed the relative accelerations too. This suggests looking again for a relative-acceleration-based formulation.

\citeN{SaundersPrincipia} has provided an elegant formulation along these lines, albeit with a rather different motivation. His theory (which I call here \emph{vector relationism}) is most simply expressed starting from the discrete form (\ref{newton-sum}) of Newton's force law. Saunders observes that the equations can be rearranged into the $N(N-1)$ equations for relative acceleration
\be \label{difference}
\ddot{x}_m^i(t)-\ddot{x}_n^i(t)=
\sum_{k \neq m}G m_k \frac{(x^i_k(t)-x^i_m(t))}{|x_k(t)-x_n(t)|^3} 
-
\sum_{k\neq n}G m_k \frac{(x^i_k(t)-x^i_n(t))}{|x_k(t)-x_m(t)|^3} 
\ee
together with an equation for the centre of mass,
\be \label{COM}
\dbdbd{}{t}\left(\frac{\sum_n x^i_n(t)m_n}{\sum_n m_n} \right)=0.
\ee
But the centre of mass is unobservable (at least when considering genuinely isolated systems, and in particular for the Universe as a whole) and all of the physics is contained within the difference equations (\ref{difference}). These equations remain well-defined even in the cosmological context: they may be resolved into a two-body interaction plus a sum of tidal effects on those two bodies from the other bodies, and the latter decrease with $|x|^3$ at large distances. So Saunders proposes simply dropping (\ref{COM}) and taking (\ref{difference}) to define Newtonian gravity (he argues, in fact, that Newton himself must be read as tacitly adopting this approach in order to make sense of the Principia). 

I discuss Saunders' vector relationism in detail in \citeN{wallacenewton} (see also \citeN{weatherallnewtoncartan} for discussion of its relation to Newton-Cartan gravity). For our purposes, the important thing about it is that it provides unambiguous predictions for the relative accelerations, and so is not plagued by the indeterminism of the potential theory.

To clarify the comparison with that theory, let's obtain the continuum limit of vector relationism. The relative accelerations of two test particles at positions $x$ and $y$ is given by
\be 
\dot{v}^i(x,t)-\dot{v}^i(y,t)=\int \dr{x}'^3 \rho(x',t) (F^i(x-x')-F^i(y-x'))
\ee
where 
\be 
F^i(x)=G\frac{x^i}{|x|^3}.
\ee
So the relative acceleration of infinitesimally close test particles is given by
\be 
\nabla_j\dot{v}^i(x,t)=\int \dr{x}'^3 \rho(x',t) \nabla_j F^i(x-x')
\ee
Inserting a factor $\e{-\epsilon/|x|}$ into $F^i(x)$ to regularise the singularity, differentiating, and taking $\epsilon \rightarrow 0$, we obtain
\be 
\nabla_j F^i(x) = 4 \pi G\delta(x) \delta^i_j+ \frac{G}{|x|^5}\left(3 x^i x^j - |x|^2 \delta^{ij}\right)
\equiv 4 \pi G\delta(x)\delta^i_j + G D^{ij}(x)
\ee
so that the relative acceleration is
\be \label{relac-saunders}
\nabla_j\dot{v}^i(x,t)=\frac{4}{3}\pi \rho(x,t)\delta^i_j+\int \dr{x}'^3 \rho(x',t) D^{ij}(x-x').
\ee
If we compare (\ref{relac-saunders}) with (\ref{relac-potential}), we recover Poisson's equation for the scalar curvature, but in addition we have a closed-form expression for the Newtonian Weyl tensor,
\be \label{weyl}
E_{ij}(x,t)=G\int \dr{x'}^3 \rho(x') \frac{3(x_i-x'_i)(x_j-x'_j) -|x-x'|^2 \delta_{ij}}{|x-x'|^5}.
\ee
Unlike the scalar curvature, the expression for the Weyl tensor is highly nonlocal, something to be expected given the action-at-a-distance theory with which we began.

With this expression in hand, we can obtain a potential-theory boundary condition that reproduces vector relationism. To do so, firstly define the average density by
\be 
\bar{\rho}(t)=\lim_{r\rightarrow \infty}\frac{1}{(4/3)\pi r^3}\int^{|x|<r}\dr{x}^3 \rho(x,t)
\ee
assuming that it exists (if it doesn't, even the relative-acceleration version of Newtonian gravity is likely to be ill-defined). We can then write $\rho(x,t)=\bar{\rho}(t)+\delta \rho(x,t)$ and define
\be \label{Spotential}
\Phi(x,t)=\frac{4}{3}\pi G \bar{\rho}(t) |x|^2 + G\int\dr{x'}^3 \frac{\delta \rho(x',t)}{|x-x'|}.
\ee
For reasonably behaved deviations $\delta \rho$ from homogeneity (including as a special case `island universes' where $\bar{\rho}=0$), this expression can be expected to converge, since the average of $\delta \rho$ over successively larger regions of space tends to zero. Rigorous establishment of the convergence conditions lies beyond the scope of this paper; I will assume some condition on $\delta \rho$ such that the integral converges, and converges to some function strictly slower-growing than $x^2$, so that the first term in (\ref{Spotential}) dominates for large $|x|$.

 We can readily confirm that
\be 
\nabla_i\nabla_j \Phi(x,t)=\frac{4}{3}\pi G (\bar{\rho}(t)+\delta \rho(x,t))\delta_{ij}+E_{ij}(x,t).
\ee
In other words, $\Phi$ is a potential for the Newtonian geodesic deviation, satisfying Poisson's equation. It is unique up to time-dependent linear terms (these correspond to the time-dependent translation symmetry of the theory).

Now suppose we impose the following boundary condition on solutions $\Phi$ to Poisson's equation:
\be \label{newboundary}
\lim_{|x|\rightarrow \infty} \frac{\Phi(x,t)}{|x|^2}= \frac{4}{3}\pi G \bar{\rho}(t).
\ee 
By comparison with (\ref{gen-sol}), we can see that two solutions satisfying this condition differ by at most an unobservable linear term. Given our assumptions, (\ref{Spotential}) does satisfy the condition, so any solution that satisfies it is equal to that potential up to a linear term. We can then restate vector relationism as a boundary condition (\ref{newboundary}) on the standard potential-based version of the theory. (Note that this condition serves \emph{inter alia} to justify the perturbative expression (\ref{perturb}).) However, the ultimate justification for what might otherwise seem a somewhat ad hoc condition is the closed-form expression already obtained for the Newtonian relative acceleration.

In particular, vector relationism gives a concrete value for the Weyl tensor in homogeneous but non-isotropic universes. Namely, since (\ref{weyl}) clearly\footnote{If it's not clear, note that a constant potential transforms like a scalar under rotations, whereas the kernel of (\ref{weyl}) transforms like a traceless matrix; these are orthogonal, so under integration their product vanishes. (Or just evaluate the integral by brute force; \`{a} chacun son go\^{u}t.)} vanishes when $\rho$ is constant (and does not depend on the velocity field, isotropic or otherwise) then in homogeneous universes, $E^i_j=0$. 

In fact, this condition on the Weyl tensor for homogeneous and non-isotropic universes has been (in effect) proposed before in previous force-based versions of Newtonian cosmology, notably by \citeN{narlikar1963newtonian} and Davidson and Evans (\citeNP{davidson1973newtonian,evans1974newtonian}). Those previous versions were developed in a somewhat heuristic way relying on physical intuition; Saunders' vector relationism can be usefully considered as a rigorous framework from which these previous versions can be derived.

So: vector relationism replaces the pernicious indeterminism of the potential theory with  unambiguous predictions for the Weyl tensor, and in doing so lets us derive a clean and physically motivated boundary condition for the potential. And it does it all in a way that's very true to the original structure of Newtonian gravity. What's not to like?

Just this: those `unambiguous predictions' are unambiguously in conflict with general relativity.

\section{Anisotropic cosmologies and their Newtonian limits}

The simplest anisotropic homogenous spacetime --- the `Bianchi type I spacetime' --- has metric
\be 
\dr{s}^2 = \dr{t}^2 - \sum_i A_i^2(t) \dr{X}^i \dr{X}^i.
\ee
(In this section I suppress the summation convention.) Comparing this to Newtonian cosmology is awkward because the Bianchi spatial coordinate expands with the spacetime whereas the Newtonian spatial coordinate measures fixed spatial distance (in the terminology of fluid dynamics, the former uses Lagrangian rather than Eulerian coordinates). But by defining new spatial coordinates $x^i=A_i(t)X^i$, we can rewrite the Bianchi metric as
\be 
\dr{s}^2 = (1 -2 \Phi)\dr{t}^2  - \sum_i \dr{x}^i \dr{x}^i - 2 \sum_i F_i x_i \dr{t}\dr{x}^i
\ee
where 
\be 
\Phi(x,t) = \frac{1}{2}\sum_i (\dot{A}_i/A_i)^2 (x^i)^2; \,\,\,\, F_i=\dot{A}_i/A_i.
\ee
We now assume that $A_i$ varies slowly, so that $\ddt(\dot{A}_i/A_i) \ll \dot{A}_i/A_i \ll 1$, and we restrict our attention to volumes such that the relative speed of comoving particles at different points in the region are $\ll 1$, and to test particles whose speed relative to a comoving particle is $\ll 1$. Under these assumptions the contribution of $F_i$ to the geodesic equation becomes negligible compared to that of $\Phi$ and we can approximate the metric by
\be 
\dr{s}^2 = (1 -2 \Phi)\dr{t}^2  - \sum_i \dr{x}^i \dr{x}^i.
\ee
This is the familiar Newtonian limit of general relativity (cf, \egc, \citeN[pp.74--90]{waldrelativitybook}), with $\Phi$ as the Newtonian potential. But $\Phi$ is not the isotropic potential predicted by vector relationism: it has an anisotropic component,
\be \label{weyl-bianchi}
E^i_j = \delta^i_j (\dot{A}_i/A_i) - \delta^i_j\frac{1}{3}\sum_k(\dot{A}_k/A_k).
\ee

This conflict between the vanishing-Weyl-tensor prediction of force-based approaches to Newtonian cosmology, and the nonvanishing Weyl tensor of the Bianchi type I spacetime, has long been known in theoretical cosmology (see, \egc, \citeN{zeldovich1965anisotropic}, \citeN{szekeres1977}, \citeN{ellis1997newtonian}). The relative-acceleration reformulation of Newtonian physics provides a conceptually rigorous way to derive that prediction but in no way resolves the conflict. In anistropic cosmologies, there is a well-defined Newtonian limit, but it is a potential-theory limit with a nontrivial Weyl tensor, and not vector relationism.

\section{Conclusion}

There is nothing mysterious about boundary conditions in a physical theory, where that theory is intended to describe subsystems of a larger universe and the effect of that larger universe on the subsystem can be idealised to a fixed background form. For instance, a system falling freely, and sufficiently small that tidal forces across its length can be neglected, might be idealised as a system with $\Phi=$constant at infinity. A system stationary on the surface of the Earth might be idealised as having boundary condition
\be 
\lim_{|x|\rightarrow \infty} \Phi(x) = b_i x^i,
\ee
representing the fact that the system is accelerating relative to the local inertial frames. A system moving inertially, \emph{not} small enough to neglect tidal forces, but small enough to neglect the variations in those tidal forces, might be idealised as having boundary condition
\be 
\lim_{|x|\rightarrow \infty} \Phi(x) = \Phi_0 (2z^2-x^2-y^2)
\ee
(this is the second-order term in the expansion of a $1/r$ potential, after a constant term which can be discarded and a linear term corresponding to uniform free fall).

Newtonian ``cosmology'', as used in practice, can be consistently viewed the same way. Given a region that is large and homogeneous --- but small compared to the horizon size --- in a larger cosmology treated via general relativity, we can read off a boundary condition for that region from the weak-field limit of the cosmology and idealise it as a condition at infinity. So the Newtonian physics of a patch in an isotropic universe has boundary condition $E^i_j=0$ at infinity. An anisotropic universe instead requires (\ref{weyl-bianchi}) as a boundary condition.

But a cosmological theory  in the full sense is a theory not of a subsystem with external boundary, but of the Universe as a whole. Such a system cannot have boundary conditions imposed from without via considerations of its environment.

Standard potential-based Newtonian cosmology is not suited as such a theory: the need for boundary conditions specified independently at each instant of time renders it radically indeterministic. Saunders' reformulation of Newtonian gravity in terms of relative acceleration is \emph{conceptually} suited to act as a cosmology, but fails to match the predictions of general relativity away from the exactly-isotropic case. 

Put another way: the move from relative-acceleration-based to potential-based formulations of Newtonian gravity is not simply a reformulation of the same theory. It genuinely increases the modelling capacity of the theory, and that increase has cosmological applications. But it comes at a price: we have to include boundary conditions derived from the theory's embedding in some larger theory.

I conclude, tentatively, that there is no fully satisfactory Newtonian cosmology. It is not that we lack any version of Newtonian gravity that remains well-defined in cosmological contexts; indeed, we have two to choose from. One of them is a self-contained cosmological theory --- if Newton had constructed a cosmology, surely it would have been this --- but fails to reproduce the full nonrelativistic limit of general relativity. The other reproduces that limit just fine, but cannot in the full sense be regarded as cosmological.

\section*{Acknowledgements}

This paper benefitted from useful feedback by Erik Curiel and Eleanor Knox.

\end{document}